\documentclass{aastex}
\usepackage{graphicx}

\begin{document}

\title{Wide radio beams from $\gamma$-ray pulsars}
\author{V. Ravi, R. N. Manchester, G. Hobbs}
\affil{Australia Telescope National Facility, CSIRO, Epping NSW 1710, Australia}

\begin{abstract}

  We investigate the radio and $\gamma$-ray beaming properties of
  normal and millisecond pulsars by selecting two samples from the
  known populations. The first, Sample G, contains pulsars which are
  detectable in blind searches of $\gamma$-ray data from the
  \emph{Fermi} Large Area Telescope. The second, Sample R, contains
  pulsars detectable in blind radio searches which have spin-down
  luminosities $\dot{E} > 10^{34}$~erg s$^{-1}$. We analyse the
  fraction of the $\gamma$-ray-selected Sample G which have detectable
  radio pulses and the fraction of the radio-selected Sample R which
  have detectable $\gamma$-ray pulses. Twenty of our 35
  Sample G pulsars have already observed radio pulses. This rules out
  low-altitude polar-cap beaming models if, as is currently believed,
  $\gamma$-ray beams are generated in the outer magnetosphere and are
  very wide. We further find that, for the highest-$\dot{E}$ pulsars,
  the radio and $\gamma$-ray beams have comparable beaming factors,
  i.e., the beams cover similar regions of the sky as the star
  rotates. For lower-$\dot{E}$ $\gamma$-ray emitting pulsars, the
  radio beams have about half of the $\gamma$-ray sky coverage. These
  results suggest that, for high-$\dot{E}$ young and millisecond
  pulsars, the radio emission originates in wide beams from regions
  high in the pulsar magnetosphere, probably close to the null-charge
  surface and to the $\gamma$-ray emitting regions. Furthermore, it
  suggests that for these high-$\dot E$ pulsars, as in the
  $\gamma$-ray case, features in the radio profile represent caustics
  in the emission beam pattern. 

\end{abstract}


\section{Introduction}

The observed pulses from pulsars are interpreted as an emission beam,
fixed relative to the rotating neutron star, sweeping across the
Earth. The magnetospheric emission-region structure, the viewing angle
with respect to the rotation axis and the angle between the magnetic
and rotation axes define the shape of the observed pulse
profile. Pulsar radio emission is generally thought to originate
within a few percent of the light-cylinder radius ($R_{\rm LC}$) above the polar caps
\citep[e.g.][]{lm88}.  Many pulsars exhibit a ``radius-to-frequency''
mapping \citep{cor78} which is indicative of polar cap emission. The
increasing pulse width at lower radio frequencies is
interpreted as emission at increasing heights above the poles along
more widely directed field lines.  The radio profiles of many young
pulsars and millisecond pulsars (MSPs) have an ``interpulse'', a
second pulse component separated from the main pulse by close to
$180\degr$ of pulse phase, or widely separated pulse components. In
most cases, such pulsars do not exhibit radius-to-frequency
mapping, that is, they have frequency-independent pulse component separations
\citep{hf86}. 

The known population of $\gamma$-ray pulsars has recently been greatly
expanded by the six-month data release of the \emph{Fermi} Large Area
Telescope (LAT) \citep[][hereafter A10]{aaa+10c}. This sample of 46
$\gamma$-ray pulsars includes both young pulsars and MSPs, all with
high spin-down luminosities:
\begin{equation} 
     \dot{E}=\frac{4\pi^{2}I\dot{P}}{P^{3}}>10^{33}\,{\rm erg\,s^{-1}},
\end{equation}
where $I=10^{45}$\,gm\,cm$^{-2}$ is the assumed neutron star moment of
inertia, $P$ is the rotation period in seconds and $\dot{P}$ is the
period time-derivative. 

Recent studies indicate that $\gamma$-ray emission from pulsars
originates at high altitudes in the pulsar magnetosphere, near
$R_{\rm LC}$ and from the last open field lines
\citep[e.g.,][A10]{aaa+09,aaa+09d}. Two classes of high-altitude
particle acceleration and emission model are generally considered:
slot-gap accelerators \citep{mh04a,dr03} from the polar cap outwards
along the last open field lines, and outer-gap accelerators
\citep{ry95,cz98} where the last closed field lines cross the
null-charge surface defined by ${\bf \Omega\cdot B} = 0$ ($\bf \Omega$
is the rotation velocity vector, and $\bf B$ is the magnetic field
vector). These models result in what we shall term \emph{wide-beam}
emission geometries.  Low-altitude models for $\gamma$-ray emission
have been largely ruled out by the generally wide and multi-component
$\gamma$-ray pulse profiles and the existence of pulsed $\gamma$-ray
emission at energies greater than a few GeV; at low altitudes such
emission is precluded by electron-positron pair
absorption. Furthermore, \citet{rw09} find that outer-gap models are
statistically preferred over slot-gap models based on fits to
LAT pulse profiles for a sample of bright $\gamma$-ray
pulsars.

Four of the five brightest $\gamma$-ray pulsars detected by the
Energetic Gamma-Ray Experiment Telescope (EGRET) are also radio
pulsars \citep{tho08}. This, together with the similar radio
and $\gamma$-ray pulse morphology for many young pulsars and MSPs, led
\citet{man05} to suggest that, for these high-$\dot{E}$ pulsars, the
radio emission is also emitted in wide beams from the outer
magnetosphere near $R_{\rm LC}$. Within the framework of polar cap
emission, \citet{jw06} and \citet[][hereafter KJ07]{kj07} proposed
that the radio emission from young pulsars is emitted at intermediate
altitudes of $\sim1000$\,km or $\sim 0.2 R_{\rm LC}$ in order to account for
wide profiles of young pulsars.

In this letter, we identify samples of $\gamma$-ray-selected and
radio-selected pulsars and consider the pulsars in each sample that
also emit radio and $\gamma$-ray pulses respectively. We then discuss
the implications of our analysis for the radio beams of young pulsars
and MSPs.

\section{The $\gamma$-ray and radio pulsar samples}
We examine two pulsar samples: a $\gamma$-ray-selected sample (Sample G) and 
a radio-selected sample (Sample R). Each sample consists of a set of 
pulsars detectable using ``blind search'' techniques in the relevant band.

\subsection{The $\gamma$-ray-selected Sample G}
A10 presented a catalogue of $\gamma$-ray pulsars detected using two
different techniques. First, 29 $\gamma$-ray pulsars were found using
timing ephemerides, mostly from contemporaneous radio observations of
218 pulsars with $\dot{E}>10^{34}$\,erg\,s$^{-1}$
\citep{sgc+08}. Twenty of these were young pulsars, and nine were
MSPs. The second method involved a blind search for
$\gamma$-ray pulsars at the positions of $\sim$300 unidentified
$\gamma$-ray sources in the six-month LAT all-sky image.  This search
was sensitive to periodicities between 0.15\,s and 2\,s, and yielded
16 $\gamma$-ray pulsar discoveries. A further two pulsars with
existing $\gamma$-ray rotation ephemerides -- Geminga and PSR
J1124$-$5916 -- were also re-detected. While Geminga is well known for
its lack of radio pulses \citep{kl99}, PSR J1124$-$5916 is a radio
pulsar for which a $\gamma$-ray rotation ephemeris was used because
the extremely weak radio emission makes radio timing difficult
\citep{cmg+02}.

Sample G consists of all $\gamma$-ray pulsars found in the LAT blind
search. We further include those pulsars found using radio ephemerides
that have phase-averaged $\gamma$-ray fluxes above the blind search
sensitivity limit.\footnote{The LAT blind search sensitivity limit is
  plotted as a function of the diffuse $\gamma$-ray background in
  Figure 10 of A10.}  This is a conservative estimate, as some of the
LAT blind search pulsars are in fact below the limit. Our inclusion in
Sample G of pulsars where the $\gamma$-ray pulses were first detected
using radio-based ephemerides is justified because the blind search
over the unidentified $\gamma$-ray point sources was carried out after
the search using radio ephemerides. In total, we included the 18
pulsars found in the $\gamma$-ray blind search and 17 pulsars found
using radio ephemerides, of which three are MSPs. Searches for radio
pulses from 15 of the blind search pulsars by \citet{crr+09} have
revealed two detections. We do not discard the three blind search
pulsars that have not been searched for radio emission by
\citet{crr+09} because their positions have already been searched by
the Greenbank low-luminosity pulsar survey \citep{dtws85} with a
limiting flux density of 2~mJy at 400~MHz . Therefore, of the 35
$\gamma$-ray pulsars in Sample G, 20 are radio pulsars.

\subsection{The radio-selected Sample R}
Sample R was formed from the 218 high-$\dot E$ radio pulsars observed
by the LAT, of which 25 are also $\gamma$-ray pulsars. $\dot{E}$
values were corrected for the Shklovskii effect \citep{shk70} using
values of $\dot{P}$, proper motion and NE2001 distances \citep{cl02}
given in the ATNF Pulsar
Catalogue\footnote{http://www.atnf.csiro.au/research/pulsar/psrcat}
\citep{mhth05}; only pulsars with corrected
$\dot{E}>10^{34}$\,erg\,s$^{-1}$ were included in Sample R. We note
that the Sample G pulsars have already had the Shklovskii corrections
applied (A10).

PSR B1509$-$58 was not detected in the LAT data, but was found by the
AGILE $\gamma$-ray telescope \citep{ppp+09} and so we include it in
Sample R. Another pulsar, PSR J0034$-$0534, was detected using 13
months of \emph{Fermi} LAT data \citep{aaa+10b}. Fourteen of the
high-$\dot{E}$ pulsars were not detected in blind radio surveys, but
were found in deep searches of supernova remnants (SNRs)
\citep[e.g.][]{crgl09} and EGRET error boxes
\citep[e.g.][]{ojk+08}. We discard these 14 pulsars from Sample
R. This leaves 200 radio-selected pulsars in Sample R, of which 17 are
$\gamma$-ray pulsars.  Table~\ref{tb:samples} summarizes the two
samples.

\section{Implications for the radio beams}

We consider the radio and $\gamma$-ray beams as being defined by
emissivity functions $F_{i}(\theta,\phi)$, where the index $i=(r,\;g)$
indicates, respectively, the radio and $\gamma$-ray bands.  The
co-latitude angle $\theta$ is measured from the rotation axis and
$\phi$ is the longitudinal angle. The ``latitude coverage'',
$C_{i}(\theta)$, of a beam is defined as the longitude integration of
$F_{i}$:
\begin{equation}
C_{i}(\theta)= \int_{0}^{2\pi} F_{i}(\theta,\phi)d\phi.
\end{equation}
The fraction of the celestial sphere swept out by each beam -- the
``beaming fraction'' -- is given by
\begin{equation}\label{eq:beamfr}
  f_{i}=\frac{1}{4\pi}\int_{0}^{\pi}\int_{0}^{2\pi} F_{i}(\theta,\phi)\sin\theta d\phi d\theta=\frac{1}{4\pi}\int_{0}^{\pi}C_{i}(\theta)\sin\theta d\theta
\end{equation}
where the normalisation is such that $0 \le f_i \le 1$. 

\begin{figure}
\epsscale{0.35}
\plotone{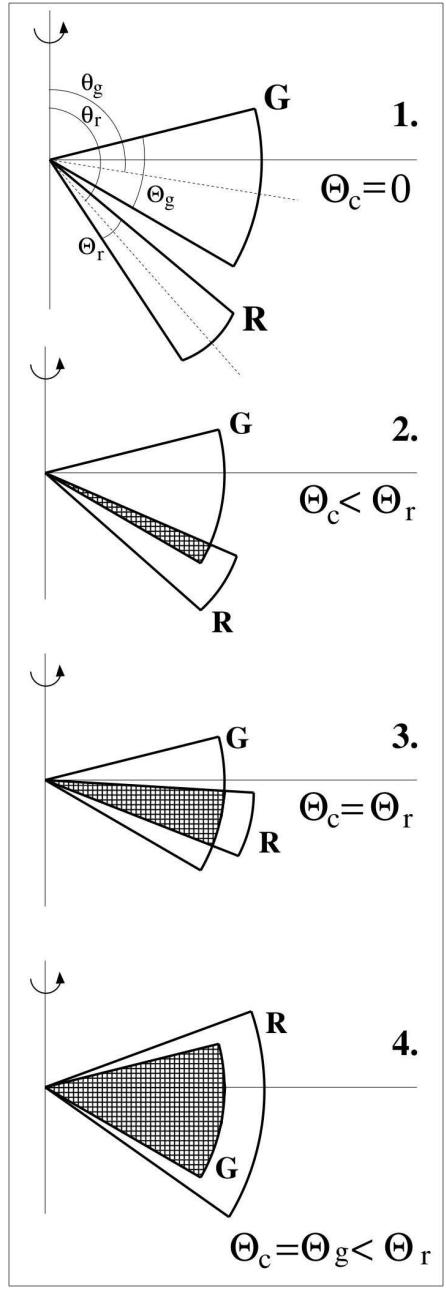}
\caption{Four possible configurations for the radio and $\gamma$-ray
  beams from a pulsar emitting in both bands. The sectors represent
  the latitude coverage of the sky as the pulsar rotates for the
  $\gamma$-ray beam (G) and the radio beam (R) where the beams are
  approximated by ``top-hat'' functions for clarity. The beams are
  centered at co-latitudes $\theta_g$ and $\theta_r$ and have
  equivalent widths of $\Theta_g$ and $\Theta_r$ for the $\gamma$-ray
  and radio bands respectively. The angular zone common to the two
  beams is shown as a hatched area and has a width $\Theta_c$. If the
  observer's line of sight lies within the hatched area, both radio
  and $\gamma$-ray pulses can be observed. }\label{fg:beams}
\end{figure}

To assist with visualisation, we approximate the beams by ``top-hat''
functions where $C_i$ is constant (and non-zero) over some beamwidth
$\Theta_i$ and zero elsewhere as illustrated in
Figure~\ref{fg:beams}. $\Theta_i$ is effectively the equivalent width
of the beam in latitude. We note that the beaming fraction $f_i$
defined by Equation~\ref{eq:beamfr} corresponds to the $f_{\Omega}$
term defined in \citet{wrwj09} for such a top-hat beam. There are four
possible configurations for a pulsar emitting in both the radio and
$\gamma$-ray bands: 1. The radio and $\gamma$-ray beams have no
overlap. 2. Partial overlap of the beams. 3. The radio beam is
narrower than the $\gamma$-ray beam and entirely overlapping. 4. The
$\gamma$-ray beam is narrower than the radio beam and entirely
overlapping.

If we assume that pulsar rotation axes are randomly oriented in space
and that pulses are detectable for observer lines of sight within the
beams, then the number of pulsars detectable in band $i$ is
proportional to the beaming fraction $f_i$. Furthermore, provided the
beam co-latitude $\theta_i$ is not too small, then the number of
pulsars detectable in band $i$ is approximately proportional to the
beamwidth $\Theta_i$. Importantly, the number detected in both bands
is proportional to the width of the overlap region $\Theta_c$.

The issue of pulsar detectability is of course critical. It is
well-known that the $\gamma$-ray luminosity of a pulsar, $L_g$, is
strongly dependent on its spin-down luminosity $\dot{E}$.  While some
$\gamma$-ray emission models predict $L_g\propto\dot{E}^{\frac{1}{2}}$
\citep[e.g.,][]{har81a}, we assume an empirical relation
$L_g\propto\dot{E}$ after \citet{sgc+08}. This implies the commonly
used $\gamma$-ray detectability metric of the pulsar spin-down flux at
the telescope, $\dot{E}d^{-2}$, where $d$ is the pulsar distance. In
contrast, the radio luminosity of pulsars is essentially uncorrelated
with both $\dot E$ and $\dot{E}d^{-2}$. Given that nearly 2000 radio
pulsars are known compared to just 46 $\gamma$-ray pulsars, it is
reasonable to assume that selection effects are much more significant
in the $\gamma$-ray band and that all $\gamma$-ray pulsars are
emitting potentially detectable radio beams.

Therefore, for the $\gamma$-ray-selected Sample G, the radio fraction,
i.e., the ratio of the number of radio detections to the total,
${\cal F_{\rm r,G}} = 20/35$, is an estimate of the ratio $\Theta_c/\Theta_g$. For
configuration 3 in Figure~\ref{fg:beams}, $\Theta_c = \Theta_r$; in
all other configurations $\Theta_c < \Theta_r$. Therefore,
\begin{displaymath}
\frac{f_{r}}{f_g} \approx \frac{\Theta_r}{\Theta_g} \geq {\cal F_{\rm r,G}}\approx0.571.
\end{displaymath}
Since the values of $\cal F_{\rm r,G}$ follow a binomial distribution, the
standard error on this mean value is
\begin{displaymath}
  \sigma = \sqrt{\frac{{\cal F_{\rm r,G}} (1-{\cal F_{\rm r,G}})}{n}} = 0.084,
\end{displaymath}
where $n=35$ is the number of pulsars in the sample. 

We conclude that, on average for $\gamma$-ray emitting
pulsars, the radio beaming fraction is at least half the $\gamma$-ray
beaming fraction. It is possible that some of the current
Sample G pulsars will be detected in the radio band,
which would increase ${\cal F_{\rm r,G}}$. With the continued
operation of the LAT, it is certain that the sample of
$\gamma$-ray-selected pulsars will increase. The value of ${\cal
  F_{\rm r,G}}$ may either increase or decrease, but its uncertainty
will decrease. The longitudinal geometries of the beams are not
constrained; therefore the radio and $\gamma$-ray emission regions
need not be co-located.

To quantify the true latitude coverage of the radio beams with respect
to the $\gamma$-ray beams, as opposed to just constraining the
overlapping area, we compare the radio-selected Sample R with Sample
G. First, given that the LAT is only sensitive to pulsars with high
values of $\dot{E}d^{-2}$, the $\gamma$-ray fraction in Sample R,
${\cal F_{\rm g,R}} = 17/200$, is not significant to this
discussion. Instead, we order the pulsars in the two samples by
decreasing $\dot{E}$ and consider the ratios ${\cal F_{\rm r,G}}$ and
${\cal F_{\rm g,R}}$ as functions of $\dot{E}$. For each pulsar, we
compute the ratios for the sub-sample of pulsars with $\dot{E}$
greater than or equal to the value for that pulsar.

\begin{figure*}
\epsscale{0.8}
\plotone{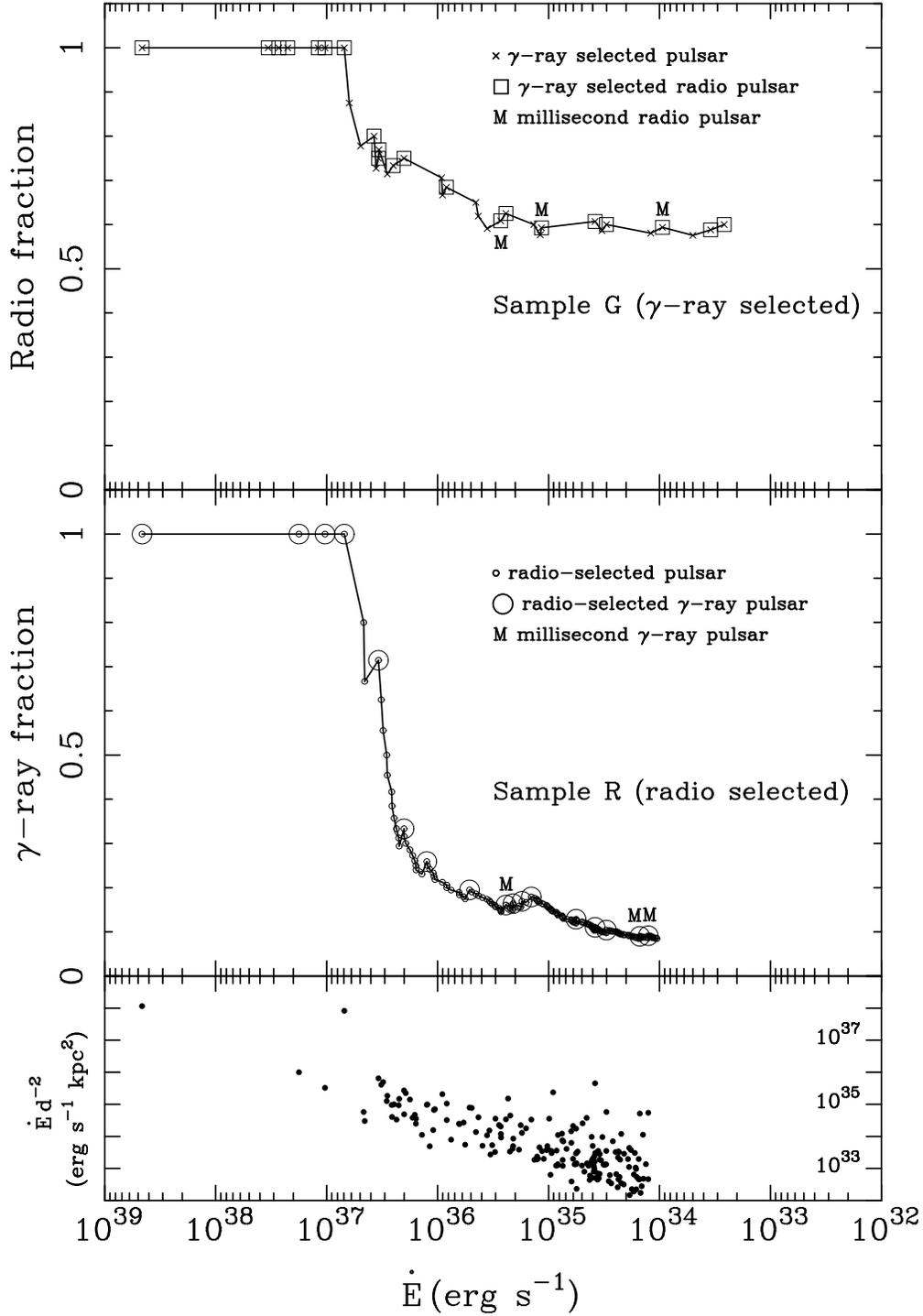}
\caption{The radio-detected fraction of Sample G, ${\cal F_{\rm r,G}}$
  (top), and the $\gamma$-ray detected fraction of Sample R, ${\cal
    F_{\rm g,R}}$ (middle), computed for each pulsar for the sub-sample
  of all pulsars having $\dot{E}$ greater than or equal to the value
  for that pulsar. Note that $\dot{E}$ increases to the left on the
  plot.  The bottom plot shows values of $\dot{E}d^{-2}$ for the
  Sample R pulsars.}\label{fg:samp}
\end{figure*}

The results are shown graphically in Figure~\ref{fg:samp}. It is
striking that the seven highest-$\dot{E}$ Sample G pulsars are
detected in both bands. Within the same $\dot{E}$ range, all four
Sample R pulsars are also $\gamma$-ray emitters. It is therefore
evident that ${\cal F_{\rm r,G}}$ and ${\cal F_{\rm g,R}}$ have
comparable values in this high-$\dot{E}$ regime. For lower-$\dot{E}$
values and correspondingly lower-$\dot{E}d^{-2}$ values in Sample R
(shown in the bottom panel of Figure~\ref{fg:samp}) ${\cal F_{\rm
    g,R}}$ reduces sharply, but this is entirely due the
$\dot{E}d^{-2}$ selection for $\gamma$-ray pulsar detections.
However, it is significant that some radio-only pulsars with $\dot{E}$
between $2\times10^{36}$ and $4\times10^{36}$~erg\,s$^{-1}$ have high
values of $\dot{E}d^{-2}$. These pulsars probably have $\gamma$-ray
beams not directed toward us, suggesting that the radio beams for
these high-$\dot{E}$ pulsars are of comparable width to the
$\gamma$-ray beams and do not always overlap with their latitude
coverage. Any of configurations 1, 2, or 4 in Figure~\ref{fg:beams}
could be applicable to this case.

In contrast, under the assumption that radio pulse detectability is
independent of $\dot{E}$, Sample G is free of these selection effects
and the drop in ${\cal F_{\rm r,G}}$ at
$\dot{E}\sim6\times10^{36}$\,erg\,s$^{-1}$ can be interpreted as an
evolution in ${\cal F_{\rm r,G}}$ from unity for high-$\dot{E}$
$\gamma$-ray pulsars to $\sim0.45$ for lower-$\dot{E}$ $\gamma$-ray
pulsars. In general, the highest-$\dot{E}$ pulsars are among the
youngest known pulsars, as determined, for example, by SNR
associations \citep{kas00}. Furthermore, pulsars evolve toward lower
$\dot{E}$ values. It therefore appears that, for the highest-$\dot{E}$
pulsars, the radio and $\gamma$-ray beams have comparable
beaming fractions with largely overlapping radio and $\gamma$-ray
beams. As these pulsars age, the radio beams evolve to overlap about
half the latitude coverage of the $\gamma$-ray beams. Ultimately the
pulsars become undetectable in the $\gamma$-ray band and the radio
beams become even narrower as they evolve toward the beaming fractions of
$\sim 0.1$ typical for older pulsars \citep{tm98}.

\section{Millisecond pulsars}
The lower-$\dot{E}$ pulsars in Samples G and R include five
MSPs,\footnote{While \citet{aaa+09f} report the $\gamma$-ray
  detections of eight MSPs, three did not fall within either of the
  cutoffs applied to Samples G or R.} marked with the letter `M' in
Figure~\ref{fg:samp}.  Despite the differences between normal pulsars
and MSPs (e.g., surface magnetic field strength and rotation period)
we do not isolate MSPs from our samples for three reasons. First, the
three MSPs in Sample G were $\gamma$-ray-selected, which implies some
commonalities with young $\gamma$-ray pulsars.  Second, the nine MSPs
detected with the LAT by \citet{aaa+09f} and \citet{aaa+10b} have
values of $\dot{E}d^{-2}$ that are among those of the top 18 MSPs
searched, which gives a $\gamma$-ray fraction of 0.5, consistent with
the lower bound on the radio fraction for the lower-$\dot{E}$ Sample G
pulsars ($\sim 0.45$). Third, it has been suggested by \citet{man05},
based on pulse profile and giant pulse characteristics, that
millisecond pulsars have wide radio beams similar to those of young
pulsars. Therefore, in the discussion below on the radio beams of
$\gamma$-ray emitting pulsars, we treat young pulsars and MSPs
together.

\section{Discussion}

To test the consistency of our results with various empirical
polar-cap models for pulsar beaming, we carried out a series of Monte
Carlo simulations for the Sample G pulsars to calculate their values
of ${\cal F_{\rm r,G}}$.  Sample R was not considered because of the
$\dot{E}d^{-2}$ selection effects discussed above. We assume that the
$\gamma$-ray beams sweep the entire celestial sphere, that is, we
assume $f_g=1$, consistent with outer-magnetosphere models
\citep[e.g.,][]{wrwj09}. For each pulsar, $10^{5}$ pairs of $\alpha$
and $\zeta$ were randomly chosen from sinusoidal distributions peaked
at $\pi/2$ ($\alpha$ is the angle between the rotation and magnetic
axes and $\zeta$ is the angle between the rotation axis and the line
of sight). The radio beam opening half-angle $\rho$, measured from the
magnetic axis, was calculated for each Sample G pulsar and beaming
model. The beaming fraction $f_r$ was then derived by finding the
fraction of points with $|\alpha-\zeta|<\rho$. Since both $\alpha$ and
$\zeta$ ranged between 0 and $\pi/2$, emission from both magnetic
poles was accounted for.

We first consider the model of \citet{ran93} with $\rho=5.8P^{-1/2}$
for which we obtain an average beaming fraction of $f_r=0.46$ for the
Sample G pulsars, somewhat lower than our estimate of $f_r/f_g
\approx f_r \geq 0.57$. We note that our assumption of $f_g=1$ is
conservative in this context; other similar simulations by
\citet{ry95} and \citet{hgg07} using theoretical predictions of $f_g$
based on the outer-gap model find $f_r =0.3$ and $f_r=0.1$
respectively, much less than the observed value. 

We then considered the model of KJ07 which, for young pulsars, has a
radio emission height above the polar caps of $\sim 1000$~km, along
with beaming along the last open field lines. This model results in
$f_r=0.74$, larger than our lower bound of $\sim 0.57$. Hence, the KJ07
radio beaming model is consistent with our results. However, the fact that
the seven highest-$\dot{E}$ pulsars in Sample G are all observed to
emit radio pulses (i.e., have $f_r \sim 1$) is not predicted by the
KJ07 model. The product of the radio beaming fractions derived from
the KJ07 model of these seven pulsars is 0.22; hence, the KJ07 model
has a 22\% chance of reproducing this result.

In contrast to the KJ07 model, outer-magnetosphere wide-beam models
for the radio emission predict $f_r\sim 1$, consistent with our result
that ${\cal F_{\rm r,G}}$ is near unity for the highest-$\dot{E}$
pulsars.  Besides this, there are other compelling reasons to consider
wide-beam models for these high-$\dot{E}$ pulsars
\citep[cf.,][]{man05}.  First, there is an excess of pulsars with
widely separated pulse components or interpulses among young,
high-$\dot{E}$ pulsars \citep[e.g., see Fig. 1 of][]{wj08}. Many of
these pulsars have pulse profiles very similar to observed
$\gamma$-ray profiles, with bridge emission between the main
components. Furthermore, they have frequency-independent component
separations, inconsistent with low-altitude polar-cap models. As was
first pointed out by \citet{nv83}, polarization results from many
young pulsars indicate beams elongated in the latitude direction, with
impact parameters, $|\alpha-\zeta|$, that are much greater than
observed pulse widths. This is inconsistent with polar-cap models but
fully consistent with outer-magnetosphere models.

We therefore propose that the radio emission region for young
high-$\dot{E}$ pulsars is located high in the pulsar magnetosphere,
relatively close to the null-charge surface and to the
$\gamma$-ray-emitting region. The evolution in ${\cal F_{\rm r,G}}$
with $\dot{E}$ suggests a slot-gap model for the radio
emission and hence that radio profile components represent caustics
just as for the $\gamma$-ray profiles. This interpretation is also
consistent with a transition to low-altitude polar-cap emission for
older pulsars. 

Not all $\gamma$-ray pulsars are seen in the radio and not all
high-$\dot{E}d^{-2}$ radio pulsars are seen in $\gamma$-rays. This
implies that, while the radio and $\gamma$-ray emission regions may be
in the same general region of the pulsar magnetosphere, they are not
necessarily co-located. In the case of PSR J0034$-$0534 however, the
radio and $\gamma$-ray profiles are almost identical and aligned in
phase, suggesting that the emission regions are in fact co-located
\citep{aaa+10b}.

\section{Conclusions}

By comparing the relative occurrence of radio and $\gamma$-ray pulsed
emission in $\gamma$-ray-selected and radio-selected samples, we find
that the radio beaming fraction $f_r$ appears to be close to unity for
the highest-$\dot{E}$ pulsars, decreasing to $\sim0.5$ for
lower-$\dot{E}$ $\gamma$-ray pulsars. Our estimated lower bound $f_r$,
averaged over all $\gamma$-ray-selected pulsars, of $\sim 0.57$ is
inconsistent with low-altitude polar-cap models for the radio emission but
is consistent with KJ07's intermediate-altitude polar-cap
model. However, the KJ07 model is marginally inconsistent with our
result that $f_r \sim 1$ for the highest-$\dot{E}$ pulsars. With other
evidence from radio and $\gamma$-ray pulse morphologies, these results
suggest that, for high-$\dot{E}$ pulsars, the radio emission
originates in wide beams from the vicinity of the null-charge surface,
possibly with a slot-gap configuration, with profile components
representing caustics in the emission pattern.

Improved statistics from further $\gamma$-ray and radio discoveries
will help to test our conclusions. Modelling of radio pulse profiles
with slot-gap or other wide-beam emission geometries for both young
high-$\dot{E}$ pulsars and MSPs would be valuable. If
outer-magnetoshpere wide-beam radio emission from these pulsars is
confirmed, it would have profound implications for pulsar population
and evolution studies as well as our understanding of radio and
$\gamma$-ray pulse emission mechanisms.

\acknowledgements

We thank Simon Johnston and Kyle Watters for valuable discussions and
an anonymous referee for constructive suggestions. GH
is supported by an Australian Research Council QEII Fellowship
(project \#DP0878388).

%\bibliographystyle{apj}
%\bibliography{journals,modrefs,psrrefs,crossrefs}

\clearpage

\begin{deluxetable}{lcc}
\tablecaption{The $\gamma$-ray-selected and radio-selected samples\label{tb:samples}}
\tablehead{\colhead{Sample} & \colhead{$\gamma$-ray pulsars} &
  \colhead{Radio pulsars}}
\startdata
$\gamma$-ray-selected (G) & 35 & \phn20 \\
Radio-selected (R) & 17 & 200 \\
\enddata
\end{deluxetable}

\end{document}